\documentclass[aps,prd,superscriptaddress,showpacs,amsmath,amssymb]{revtex4}
\usepackage{graphicx,epsf}
\usepackage[]{latexsym}
\newcommand{\be}{\begin{eqnarray}}
\newcommand{\ee}{\end{eqnarray}}

\newcommand{\Euc}{\mbox{{\rm I\hspace{-2truemm} E}}}

\newcommand{\nee}{\nonumber \end{eqnarray}}
\newcommand{\nn}{\nonumber \\}
\newcommand{\ident}{\mbox{{\rm 1\hspace{-2.8truemm} I}}}
\begin{document}
\title{PP-waves on Superbrane Backgrounds}
\author{Marco Cavagli\`{a}}
\email{cavaglia@phy.olemiss.edu }
\affiliation{Department of Physics and Astronomy,University of
Mississippi,University, MS 38677-1848, USA}
\affiliation{}
\author{Benjamin ~Harms}
\email{bharms@bama.ua.edu}
\affiliation{Department of Physics and Astronomy, The University
of Alabama, Box 870324, Tuscaloosa, AL 35487-0324, USA}
\begin{abstract}
In this paper we discuss a method of generating solutions of the
Einstein equations on supersymmetric and non-supersymmetric
backgrounds.  The method involves the embedding of a
supersymmetric spacetime into another, curved spacetime. We
present three examples with constituent spacetimes which support
``charges'', one of which was known previously and the other two
of which are new. All of the examples have PP-waves as one of the
embedding constituents.
\end{abstract}
\maketitle
\large
Supersymmetry is a unifying principle which helps overcome the divergence problem and the
``naturalness'' problem of the Standard Model \cite{susy}. All consistent string theories are
supersymmetric \cite{polchinski_book}.  Therefore the study of supersymmetric spacetimes is a
worthwhile endeavor because of the intrinsic interest of such spacetimes and because of their
importance to other theoretical investigations. There are relatively few  spacetimes which
possess at least some supersymmetries and even fewer which are maximally supersymmetric, i.e.
possess the maximum number of supersymmetries allowed by the dimension of the spacetime
\cite{Blau:2001ne}.

In this letter we present a general method for generating solutions of the Einstein equations
on supersymmetric backgrounds. The method also works for non-supersymmetric spaces and raises
the possibility of designer solutions of Einstein's equations. The application of the method
studied in the present work involves the embedding of a fractionally supersymmetric spacetime
into another spacetime plus a deformation term in the component of the metric for the coordinate which is
common to the superbranes of the constituent spacetimes.  We discuss three examples of this
method, one previously known and two which are new. The supersymmetric examples discussed here are for $D = 11$
supergravity, but the method works for supergravity in any number of dimensions.  Although the
constituents in the supersymmetric examples are plane parallel gravitational waves propagating on
supersymmetric $D = 11$ curved spacetime backgrounds, the method should work for other
backgrounds such as type IIA supergravity-superstrings. Throughout the paper capital Latin
indices run from $0$ to $10$, Greek indices from $0$ to $2$, $(m,n)=3\dots 10$, and
$(i,j)=2\dots 10$. The metric signature is $(-1,1\dots 1)$.

Gravitational waves with parallel wave fronts and parallel rays in $11$ dimensions have a
metric of the form \cite{figueroa}
\be
 ds^2 = 2\, dx^+\,
 dx^- + H(x^i,x^-)\,(dx^-)^2 + \delta_{ij}\,dx^i\, dx^j \,,
\label{ppwave}
\ee
where $x^{\pm}=x_1\pm x_0$ are light-cone coordinates, and $H(x^i,x^-)$ is a function which is
determined from the relation
\be
 \Box H = -\frac{1}{12}|\Phi|^2
\ee
in which $\Box$ is the Laplacian operator in the transverse Euclidean space $\Euc^9$, and
$\Phi$ is a 3-form in $\Euc^9$. $\Phi$ is related to the 4-form arising from the ``charge'' on
the superbrane by $F_4 = dx^- \wedge \Phi$. The Lagrangian density for this system is given by
\cite{duff}
\be
 \kappa^2{\cal L}=\frac{1}{2}\sqrt{-g}\left[R - \frac{1}{48}\,
 F_{MNPQ}F^{MNPQ}\right] + \frac{1}{2(12)^4}\varepsilon^{UVWMNOPQRST}\,
 F_{UVWM}\, F_{NOPQ}\, A_{RST} \,,
\ee
where $\varepsilon^{UVWMNOPQRST}$ is the eleven-dimensional Levi-Civita tensor density, and
$A_{MNP}$ is an antisymmetric potential with corresponding field strength given by
\be
F_{MNPQ} = 4\partial_{[M}A_{NPQ]}\,.
\label{F}
\ee
The components of the 4-form $F_4$ satisfy the field equation
\be
 \partial_M\left(\sqrt{-g}\, F^{MUVW}\right) +
 \frac{1}{1152}\varepsilon^{UVWMNOPQRST}F_{MNOP}\, F_{QRST}=0
\ee
and the metric tensor elements satisfy the Einstein equations
\be
 R_{MN} - \frac{1}{2}\, g_{MN}\, R = \kappa^2\, T_{MN}
\ee
with
\be
 \kappa^2\, T_{MN} = \frac{1}{12}\left(F_M^{\ \ PQR}\, F_{NPQR} -
 \frac{1}{8}\, g_{MN}\, F_{PQRS}\, F^{PQRS}\right)\,.
\ee
The Killing spinors for this spacetime satisfy the relation
\be
 \left(D_M - \Omega_M\right)\epsilon = 0\,,
 \label{covder}
\ee
in which $D_M$ is the covariant derivative
\be
 D_M =\partial_M + \frac{1}{4}\omega_M^{\ \ A B}\, \Gamma_{AB}\,,
\ee
and
\be
 \Omega_M = \frac{1}{288}\, F_{PQRS}\left(\Gamma^{PQRS}_{\ \
 \ \ \ \ M} + 8\, \Gamma^{PQR}\delta^S_M\right)\,.
\ee
The eleven-dimensional Dirac matrices satisfy the relation $\left\{\Gamma_A\,,\Gamma_B\right\}
= 2\,\eta_{AB}$ and for convenience we define $\Gamma_{AB..C} =
\Gamma_{[A}\Gamma_B..\Gamma_{C]}$.

The metric described in Eq.\ (\ref{ppwave}) preserves half of the maximum allowed 32
supersymmetries. For certain choices of $H(x^i,x^-)$ the space described by Eq.\
(\ref{ppwave}) is maximally supersymmetric. The latter occurs when $H(x^i,x^-)$ is of the form
\be
 H(x^i,x^-) &=& \left\{\begin{array}{ll} -\frac{1}{9}\, \mu^2\, \delta_{ij} x^i x^j
 &\qquad i,j = 2,3,4 \nn\nn
 -\frac{1}{36}\, \mu^2\, \delta_{ij} x^i x^j
 &\qquad i,j = 5,\dots, 10\end{array}\right.\nn
 \label{CW}\\
 \Phi &=& \mu\, dx^2\wedge dx^3\wedge dx^4\,,
\nonumber
\ee
where $\mu$ is a constant.  The non-zero components of the 4-form for this case are
$F_{-[234]} = \mu$, where the square brackets indicate complete antisymmetry in the indices.
Spacetimes with a null homogeneous flux which describe gravitational waves with a constant
$F_4$ are called homogeneous, plane parallel (HPP) waves. The space described by Eq.\
(\ref{CW}) is a special case of the Cahen - Wallach spaces \cite{cahen}, which are
indecomposable Lorentzian symmetric spaces.

Embedding a PP-wave in an AdS spacetime results in a supersymmetric space which typically
preserves $\frac{1}{4}$ of the maximum supersymmetry allowed by the AdS spacetime.  For
certain choices of $H(x^i,x^-)$ gravitational waves on AdS backgrounds result in so-called
``supernumerary'' supersymmetries \cite{kerimo}, which preserve $\frac{1}{2}$ the maximal
number of supersymmetries. For purely gravitational PP-waves on an AdS background the metric
is given by
\be
 ds^2 = dx_2^2 + {\rm e}^{2\, a\,x_2} \left[2\, dx^+\, dx^- +H(x^-,x_2,x_m)\, (dx^-)^2 + \delta_{mn}
 dx^m dx^n\right]\,,
\ee
where $a$ is a constant. Supersymmetric solutions to the Einstein equations with $d$-form
field strength sources for PP-waves propagating on an AdS spacetime of any dimension $D$ have
been discussed in Ref.\ \cite{kerimo}. For the case of maximal supersymmetry the Killing
spinor solution has the form
\be
 \epsilon = e^{ax_2/2}\left(1-\frac{ix_mU(x^+)\Gamma_m\Gamma_-}{2}\right)
 \left(1+\frac{ie^{-ax_2}U(x^+)\Gamma_-}{2a}\right)\left[1 - \frac{1}{2}
 \left(1-e^{-i\int U dx^+}\right) \Gamma_+\Gamma_-\right]\epsilon_0\,,
\ee
where the constant spinor $\epsilon_0$ satisfies the relation $(\Gamma_{x_2} +1)\, \epsilon_0
= 0$, and $U(x^+)$ is a function which can be determined from the integrability conditions.
Since there is only one condition on $\epsilon_0$, this spacetime preserves $\frac{1}{2}$ the
supersymmetry.

To illustrate our method of generating solutions of the Einstein equations which preserve at
least some of the supersymmetries, we consider now the embedding of the PP-wave geometry into
the superbrane solution of $D = 11$ supergravity. In Ref.\ \cite{duff} a solution of the
coupled supergravity - superbrane equations which preserves $\frac{1}{2}$ of the
supersymmetries was obtained. The metric for this space is given by
\be
 ds^2 = \left( 1 + \frac{q}{r^6}\right)^{-2/3}\,
 \eta_{\mu\nu}\, dx^{\mu}\, dx^{\nu}
 +\left(1+\frac{q}{r^6}\right)^{1/3} \delta_{mn}\, dy^m\, dy^n \,.
\label{superb}
\ee
The 4-form for the superbrane solution can be written in terms of a 3-form gauge
field as given in Eq.\ (\ref{F}) with an antisymmetric potential given by
\be
 A_{\mu\nu\rho} =\pm \frac{1}{{}^3g}\, \varepsilon_{\mu\nu\rho}
 \left(1+\frac{q}{r^6}\right)^{-1} \, ,
\ee
where $q$ is a constant, $\varepsilon_{\mu\nu\rho}$ is the three-dimensional covariant
Levi-Civita symbol, ${}^3g = {\rm det}(g_{\mu\nu})$, and $r^2 = \sum_m\, y_m^2$. The only
non-zero 3-form components are
\be
A_{[012]} = \pm \left(1+\frac{q}{r^6}\right)^{-1} \,,
\ee
for which
\be
 F_{[012]m} = \pm \frac{6\, q\, x_m}{r^8(1+q/r^6)^2}\,.
\ee
The Einstein tensor elements are
\be
 G_{\mu\nu}=(-1)^{\delta_{\mu 0}}\frac{9\, q^2\,
 r^4}{(r^6+q)^3}\delta_{\mu\nu}\,,\qquad
 G_{mn} =\left\{
 \begin{array}{ll}
  \displaystyle ~~\frac{18q^2x_mx_n}{r^4(r^6+q)^2} &\qquad m\not=n\\ \\
  \displaystyle -\frac{9q^2(r^2-2\, x_m^2)}{r^4(r^6+q)^2}&\qquad m=n\,.
 \end{array}
 \right.
\ee
These expressions were obtained for a spacetime which is invariant under ${\rm P}_3 \times
{\rm SO}(8)$, where ${\rm P}_3$ is the $D = 3$ Poincar\'{e} group. The Killing spinors can be
obtained from a covariant equation similar to Eq.\ (\ref{covder}). The Dirac matrices for this
case are
\be
 \Gamma_A=\left(\gamma_{\mu}\otimes\Gamma_9\,,
 \ident\otimes\Sigma_m\right)\,,
\ee
where $\gamma_{\mu}$ and $\Sigma_m$ are the $D=3$ and $D=8$ Dirac matrices, respectively,
and $\Gamma_9 = \Sigma_3\, \Sigma_4\dots\Sigma_{10}$. The equations for the Killing spinors
result in two nontrivial solutions
\be
 \left(1\pm \Gamma_9\right)\, \epsilon(r) = 0 \,,
\ee
where
\be
 \epsilon(r) = \left(1+q/r^6\right)^{1/6}\, \epsilon_0 \,.
\ee
Each of these solutions preserves half of the maximum possible supersymmetries. To embed the
PP-waves in the superbrane background (\ref{superb}), the line element is written as
\be
 ds^2 =\left(1 + \frac{q}{r^6}\right)^{-2/3}\left\{2\,dx^{+}dx^{-}
 -\frac{\mu^2}{2}\left[x_2^2-H(r)\right](dx^-)^2+dx_2^2\right\}
 +\left(1+\frac{q}{r^6}\right)^{1/3}\,\delta_{mn}dy^m dy^n\,.
\label{embed}
\ee
The function $H(r)$ is determined from Einstein's equations:
\be
 H(r) = C_1\ + \frac{C_2}{r^6} - \frac{q}{4\, r^4}\, ,
\label{H}
\ee
where $C_1$ and $C_2$ are constants of integration. The 4-form components are assumed to have
the same forms as those for the two constituent spaces
\be
 F_{[-+2]m} = \pm\frac{6\, q\, x_m\,}{r^8\,(1+q/r^6)^2}\,,
\qquad
 F_{-[345]} = \mu\,.
\ee
The non-zero Einstein tensor elements are
\be
\begin{array}{lll}
 G_{--} &=&
 \displaystyle -\frac{\mu^2}{8\, (r^6+q)^3}\,
 \left[4 r^6(r^6+q)^2 + 9q^3 + 36q^2x_2^2r^4 -36C_1
 q^2r^4 - \frac{36C_2q^2}{r^{2}}\right]\,,\\ \\
 G_{+-} &=&
 \displaystyle {\frac {9q^{2}r^{4}}{\left(r^{6}+q\right) ^{3}}}\,,\qquad
 G_{22} = \frac{9q^2r^{4}}{(r^6+q)^2}\,,\qquad
 G_{mn} =\left\{
 \begin{array}{ll}
  \displaystyle ~~\frac{18q^2x_mx_n}{r^4(r^6+q)^2} &\qquad m\not=n\\ \\
  \displaystyle -\frac{9q^2(r^2-2\, x_m^2)}{r^4(r^6+q)^2}&\qquad m=n.
 \end{array}
 \right.
\end{array}
\ee

Although the solution to the Einstein equations found in Eq.\ (\ref{embed}) does not preserve
any supersymmetries for $D = 11$ supergravity, it is possible that other embeddings preserving
a fraction of the maximal number of supersymmetries of the two constituent spaces may be found.
In the case of PP-waves embedded in an AdS background a solution was found which preserves all
32 of the supersymmetries \cite{figueroa}.  However in that case the background has no
``charge'' supported by a superbrane.  Whether or not our method always reduces the number of
preserved supersymmetries is a subject for future investigation.

The function $H(r)$ (Eq.\ (\ref{H})), which describes the deformation of the metric tensor
element, $g_{--}$,  for the light cone coordinate which is common to both of the charge
supporting membranes for the constituent spacetimes, arises from the embedding procedure and
is required for consistency between the Einstein tensor and the stress-energy tensor. The
deformation term is proportional to the product of the charges, $\mu$ and $q$, of the
intersecting membranes.  It distorts the wave hypersurfaces defined by the condition $x^+ =$
constant and the geodesics.

Our method works for nonsupersymmetric backgrounds as well as supersymmetric ones.  As an example
consider the metric for the $p$-brane in $D$ dimensions \cite{gregory} described by  ($D = q+p+2$)
\be
ds^2_p = R^{\Delta/(p+1)} \, \left[-dx_0^2+ \sum_{i=1}^p (dx^i)^2\right] + R^{(2-q-\Delta)/(q-1)} dr^2
+r^2\, R^{(1-\Delta)/(q-1)} \, d\Omega^2_q\,,
\ee
where
\be
\Delta = \sqrt{\frac{q(p+1)}{q+p}}\, , \hspace{5mm}{\rm and}\quad
R = \left[1-\left(\frac{r_0}{r}\right)^{q-1}\right]\,.
\ee
Embedding the PP-wave geometry into this background the metric becomes
\be
ds^2 = C(r)\left[2\, dx^+\, dx^- - {\mu^2}\bar x^2\, (dx^-)^2 +
\sum_{i=2}^{p}(dx^i)^2\right] +E(r)\, dr^2 +r^2 B(r)\, d\Omega^2_q\,,
\ee
where $\bar x^2=\sum_{i=2}^{p}(x^i)^2$ and
\be
C(r) = R^{\Delta/(p+1)}\,,\quad
B(r) = R^{(1-\Delta)/(q-1)}\,,\quad
E(r) = R^{(2-q-\Delta)/(q-1)}\,.
\ee
The non-zero 4-form components for this solution are
\be
F_{[-234]} = \left[ 2\, ( p-1)\; \mu^2 C(r)^3\right]^{1/2}
\ee
and the non-zero Einstein tensor element is
\be
G_{--} &=& - \left( p-1 \right) {\mu}^{2}\,.
\ee

The method for generating solutions to Einstein's equations described in this paper opens up
several lines of investigation. We intend to investigate the general procedure for embedding any
SUSY geometry into another SUSY geometry.  The PP-wave geometry is relatively simple and is easily
embedded into a background geometry.  We will investigate the applicability of the procedure to
more complex spacetimes. For example, we will investigate the possibility that the PP-wave
superbrane geometry obtained above can be embedded in an AdS geometry. If the latter is
successful, an interesting question is: can the resulting geometry be used to elucidate the
AdS/CFT correspondence? Another line of investigation is to study what determines how many
supersymmetries are preserved in geometries which are embedded into one another. The superbrane
geometry in Eq.\ (\ref{superb}) can be reduced to a type IIA superstring under the simultaneous
reduction of spacetime and worldvolume. The spacetime described by Eq.\ (\ref{embed}) will be
studied to see if a similar reduction can reduce the PP-wave superbrane geometry  to a
ten-dimensional superstring. In Ref.\ \cite{berenstein} the perturbative string spectrum was
obtained from the gauge theory point of view on a PP-wave.  The PP-waves incorporate the first
correction to the flat space results for certain states.  We can now investigate the question:
What spectrum is reproduced from ${\cal N}=4$ Y-M on a PP-wave superbrane geometry?
\section*{Acknowledgements}
We wish to acknowledge the collaboration of George Karatheodoris
in the initial stages of this work.  This work was supported (in
part) by U.S.\ DoE contract DE-FG05-91ER40622.

\end{document}